# Strong gravitational lensing in the Einstein-Proca theory

Guoping Li[1*]    Yan Zhang[2]    Li Zhang[1]    Zhongwen Feng[1]    Xiaotao Zu[1]

[1]School of Physical Electronics, University of Electronic Science and Technology of china,

Chengdu, Sichuan 610054, China

[2]Institute of theoretical physics, China West Normal University, Nanchong, Sichuan 637009, china

**Abstract:** Adopting the strong field limit approach, we investigate the strong gravitational lensing of a spherically symmetric spacetime in the Einstein-Proca theory. With the strong field limit coefficient, three observable quantities are obtained, which are the innermost relativistic image, the deflection angle and the ratio of the flux. Comparing the observable value and the theoretical value of the strong gravitational lensing, we can verify the effectiveness of the strong gravitational lensing model.

## 1    Introduction

Gravitational lensing is one of the predictions of Einstein's general theory relativity, and the astronomical phenomena of the gravitational lensing has been tested in the early 20$^{th}$ century[1]. According to the gravitational theory, photons could be deviated from their straight path when they pass close to a compact object. So, the phenomena of the light deflection called the gravitational lensing. The compact object casing the detectable deflection named the gravitational lens. Because the gravitational lensing is a new effective way to study the universe, it has most important status and special significance in the cosmology and the astronomy. Firstly, the gravitational lensing plays a important role of the astronomical telescope, and it can't only increase the brightness of the image which the source produced, but also can help us extract the information about the stars which are too dim to be observed. Secondly, the properties of the gravitational lens are related to the whole universe, so we can constrain the Hubble constant and the cosmological constant with the analysis of the gravitational lensing. Finally, the gravitational lensing also has some relation to the dark matter, the quasinormal modes, and the extra dimensions. In 2001, Ellis's study has proved that the magnification of the strong gravitational lensing is large, and it would be used to study the background objects which is far away form us[2]. Later on, the weak gravitational lensing has been studied, and the research shows that it has the important significance to the cosmology[3]. In 2000, Alcock's survey team verified that the micro gravitational lensing can use to realize the mass distributions of the fixed stars[4-7]. In a conclusion, the study of the gravitational lensing has become a one of the most popular research fields of the cosmology, astronomy and the theoretical physics.

In 2002, Bozza's research obtained the successful achievement in the strong gravitational lensing[8][12]. He put forward a reliable method to calculate the light deflection angle in a spherically symmetric spacetime with the Virbhadra-Ellis lens equation. The result shows that the deflection angle diverges logarithmically when the light ray is close to the strong gravitational fields. Then, Chen *et al.* studied the squashed Kaluza-Klein black hole and the squashed Kaluza-Klein Gödel black hole with this method[9-10]. Also, the Kerr black hole pierced by a cosmic string has been studied with the same way by Wei *et al.*[11]. With this method, lots of work in strong gravitational lensing has been studied, and the similar results are obtained [12-28,31]. However, the research of the gravitational lensing remained need us to continue perfecting it. At present, we expect that the gravitational lensing can play a more important role in the theoretical physics, cosmology and the astronomy, and so on.

---

[*] GuoPing Li:    li_gp2009@163.com

In this carved spacetime, Yang *et al.* has studied the Hawking radiation in the Einstein-Proca theory[29], and obtained the meaningful results. In this paper, we apply the strong gravitational limit method to this metric. Because the Proca field is a vector field, the spacetime of the Einstein-Proca theory is more general than the Reissner-Nordström spacetime, and the strong gravitational limit coefficient and the deflection angle of the Einstein-Proca spacetime are obtained with this method.

## 2 The strong gravitational lensing effect in the Einstein-Proca theory

In 1994, Torrii.T *et al.* provided a action of the Proca field in the carved spacetime[30]

$$S = \int d^2 x \sqrt{-g} \left( R + 2\Lambda - \frac{1}{4} F_{\alpha\beta} F^{\alpha\beta} + \frac{\mu^2}{2} A_\nu A^\nu \right). \tag{1}$$

Where, $F_{\alpha\beta} = \partial_\alpha A_\beta - \partial_\beta A_\alpha$ is the electromagnetic tensor, $\mu$ is the mass of the Proca particle, the metric of the action is

$$ds^2 = -N(r)^2 f(r) dt^2 + \frac{dr^2}{f(r)} + r^2 d\theta^2 + r^2 \sin^2\theta d\varphi^2. \tag{2}$$

In the spherically symmetric spacetime, Yang *et al.* has discussed the Hawking radiation in details with the Hamilton-Jacobi method, and the Hawking temperature has been obtained

$$T_{H1} = \frac{1 + \Lambda r_H^2}{4\pi r_H} \sqrt{1 - \frac{\mu^2 b^2}{4a^2} r_H^2}. \tag{3}$$

$$T_{H2} = \left( \frac{1 + \Lambda r_H^2}{4\pi r_H} - \frac{c^2 r_H^4}{64\pi r_H} \right) \sqrt{1 - \frac{\mu^2 b^2}{4a^2} r_H^2}. \tag{4}$$

Eq.(3)(4) is the expression of the Hawking temperature in the spherically symmetric spacetime. Then, we will calculate the strong gravitational lensing of this metric by Bozza's method. For the convenience, we consider the orbital plane of photons as $\theta = \pi/2$,

$$ds^2 = A(r) dt^2 - B(r) dr^2 - C(r) d\phi^2. \tag{5}$$

Here,

$$A(r) = -N(r)^2 f(r), \quad B(r) = f(r)^{-1}, \quad C(r) = r^2. \tag{6}$$

In the strong gravitational lensing, it's geometric principle tell us that when a beam of light from the source(S) pass close to the gravitational lens(L), the light ray would be deviated. In order to study the strong gravitational lensing more convenient, the situation that the source (S), the gravitational lens (L) and the observer are in the same orbital plane has been considered. According to the geometric principle of strong gravitational lensing, the relationship between the image and the source can be described by the Virbhadra-Ellis lens equation[32]

$$tan\theta - tan\beta = \frac{D_{LS}}{D_{OS}}\left[tan\theta + tan(\alpha - \theta)\right]. \tag{7}$$

Where, the parameter $\beta$ represents the angle between the source and the lensing, and $\theta$ is the angle between the image and the lensing, $\alpha$ is the deflection angle. The distance from the observer to the lensing is $D_{OL}$, and $D_{LS}$ is the distance between the source and the lensing, $D_{OS}$ represents the distance from the source to the observer. Meanwhile, two conditions of the strong gravitational lensing must be satisfied, the first one is $D_{OL} + D_{LS} = D_{OS}$, and the other one is the photon sphere equation $\frac{C'(r)}{C(r)} = \frac{A'(r)}{A(r)}$ [33-34]. The largest root($r_m$) of the photon sphere equation is the radius of photon sphere. With the angular momentum, the relationship between the impact parameter and the closest approach distance is[35]

$$u = \sqrt{\frac{C(r_0)}{A(r_0)}}. \tag{8}$$

The subscript 0 shows that the function is evaluated at $r_0$. On the basis of the geodesics equation, the angular deflection of photon in the direction of $\phi$ is[35-36]

$$\frac{d\phi}{dr} = \frac{\sqrt{B}}{\sqrt{C}\sqrt{\frac{C}{C_0}\frac{A_0}{A} - 1}}. \tag{9}$$

So, the deflection angle can be calculated as[35]

$$\alpha(r_0) = I(r_0) - \pi, \quad I(r_0) = \int_{r_0}^{\infty} \frac{2\sqrt{B}}{\sqrt{C}\sqrt{\frac{C}{C_0}\frac{A_0}{A} - 1}} dr. \tag{10}$$

Where, the A, B, C come from the Eq.(6). Obviously, the deflection angle depends on the closest approach distance directly. The $r_0$ decreasing, whereas the deflection angle increasing. When the $r_0$ decrease to a certain value, the deflection angle will exceed $2\pi$. As a result, the photon will arrive at observer after it takes a complete loop around the strong gravitational fields. On the other hand, the deflection angle will diverge if the condition $r_0 = r_m$ is satisfied. Then, the photon will be captured by the strong gravitational fields completely. In reference [12], the expression of the deflection angle in the strong gravitational fields is

$$\alpha(\theta) = -\bar{a}\ln\left(\frac{\theta D_{OL}}{u_m} - 1\right) + \bar{b}. \tag{11}$$

The impact parameter will be rewritten as

$$u_m = u\big|_{r_0 = r_m} = \sqrt{\frac{C(r_m)}{A(r_m)}}. \tag{12}$$

Where the limit coefficient of the strong gravitational fields is

$$\bar{a} = \frac{R(0, r_m)}{2\sqrt{q_m}}, \quad \bar{b} = I_R(r_m) + \bar{a}\ln\frac{2q_m}{y_m} - \pi. \tag{13}$$

In Eq.(13), the parameters $q_m, y_m$ are

$$q_m = q\big|_{r_0 = r_m}, \quad y_m = y\big|_{r_0 = r_m}. \tag{14}$$

To calculate the coefficient $(\bar{a}, \bar{b})$, we should calculate the function $f(z, r_0)$ and the integration $I_R(r_0)$ firstly

$$R(z, r_0) = \frac{2\sqrt{A(r)B(r)}}{C(r)A'(r)}(1 - A(r_0))\sqrt{C(r_0)}, \tag{15}$$

$$f(z, r_0) = \frac{1}{\sqrt{y_0 - [(1 - y_0)z + y_0]\frac{C(r_0)}{C(r)}}}, \tag{16}$$

$$f_0(z, r_0) = \frac{1}{\sqrt{pz + qz^2}}, \tag{17}$$

$$I_D(r_0) = \int_0^1 R(0, r_m) f_0(z, r_0) dz, \tag{18}$$

$$I_R(r_0) = \int_0^1 g(z, r_0) dz = \int_0^1 \left[R(z, r_0) f(z, r_0) - R(0, r_m) f_0(z, r_0)\right] dz. \tag{19}$$

Here, the parameters y、z、p、q are

$$y = A(r), \quad z = \frac{y - y_0}{1 - y_0}, \tag{20}$$

$$p = \frac{1-y_0}{C_0 A'_0}(C'_0 y_0 - C_0 A'_0), \tag{21}$$

$$q = \frac{(1-y_0)^2}{2C_0^2 A_0'^3}\left[2C_0 C'_0 A_0'^2 + (C_0 C''_0 - 2C_0'^2)y_0 A'_0 - C_0 C'_0 y_0 A''_0\right]. \tag{22}$$

According to the reference [29], Yang *et al.* have solved the metric of the spherically symmetric spacetime with the four-vector $A_\nu = \{A_0(r), 0, 0, 0\}$. Because the solution of the strong gravitational field is very difficult to solve, we expand the equation as[29]

$$f = f'(r_H)(r-r_H) + \frac{1}{2}f''(r_H)(r-r_H)^2 + O(r-r_H)^3, a = f'(r_H). \tag{23}$$

If we chose the lead order of the $f$, we will obtain two solutions with the condition that $\mu$ is very small,

$$a = \frac{1 + \Lambda r_H^2}{r_H}, \quad N(r) = 1, \tag{24}$$

$$a = \frac{16 - c^2 r_H^4 + 16\Lambda r_H^2}{16 r_H}, \quad N(r) = 1, \tag{25}$$

where $\Lambda$ is the cosmological constant, $c, r_H$ represent the uncertain constant and the horizon of the black hole respectively, and they are also related to the mass and the charge of the spacetime. Employing the lead order expression of the $f(r)$ and

$$A(r) = a(r - r_H), \quad B(r) = a(r - r_H)^{-1}, \quad C(r) = r^2. \tag{26}$$

Taking Eqs. (23), (26) into Eq.(15), the concrete form of $R(z, r_0)$ is gotten as

$$R(z, r_0) = \frac{2r_0}{ar^2}(1 - a(r_0 - r_H)). \tag{27}$$

Substituting Eqs. (23), (26) into Eqs.(16-21), and taking the approximate calculation [12], we get

$$f(z, r_0) = \frac{1}{\sqrt{a(r_0 - r_H) - \frac{ar_0^2}{r^2}(r - r_H)}}, \tag{28}$$

$$f_0(z, r_0) = \frac{1}{\sqrt{\frac{(r_0 - r_H)(1 - a(r_0 - r_H))}{r_0}z + \frac{[1 - a(r_0 - r_H)]^2 (3r_H - r_0)}{ar_0^2}z^2}}, \tag{29}$$

$$I_R(r_m) = \frac{8\sqrt{ar_H}}{(1+ar_H)^2} - \frac{4}{\sqrt{ar_H}} Log \frac{(1-ar_H) + \sqrt{(1-ar_H)^2 + 4a^2r_H^2}}{2ar_H}. \qquad (30)$$

According to Eqs .(15), (20-22), we gained some parameters of a spherically symmetric spacetime in the Einstein-Proca theory as

$$r_m = 2r_H \quad , \quad R(0, r_m) = \frac{1-ar_H}{ar_H}, \qquad (31)$$

$$q_m = \frac{(1-ar_H)^2}{4ar_H}, \quad y_m = ar_H. \qquad (32)$$

With Eq.(13), the limit coefficient of the strong gravitational lensing in the Einstein-Proca theory can be obtained

$$\bar{a} = \frac{1}{\sqrt{ar_H}}, \qquad (33)$$

$$\bar{b} = -\pi + \frac{1}{\sqrt{ar_H}} Log \frac{(1-ar_H)^2}{2a^2r_H^2} + \frac{8\sqrt{ar_H}}{(1+ar_H)^2}$$
$$- \frac{4}{\sqrt{ar_H}} Log \frac{(1-ar_H) + \sqrt{(1+ar_H)^2 + 4a^2r_H^2}}{2ar_H} \qquad (34)$$

$$u_m = \sqrt{\frac{4r_H}{a}}. \qquad (35)$$

Substituting Eqs.(33-35) into Eq.(11), the deflection angle of the strong gravitational lensing is derived

$$\alpha(\theta) = -\bar{a} Log\left(\frac{\theta D_{OL}}{u_m} - 1\right) + \bar{b}$$
$$= -\frac{1}{\sqrt{ar_H}} Log\left(\sqrt{\frac{a}{4r_H}} \times \theta \times D_{OL} - 1\right) - \pi + \frac{8\sqrt{ar_H}}{(1+ar_H)^2} \qquad (36)$$
$$+ \frac{4}{\sqrt{ar_H}} Log \frac{\sqrt{2ar_H}(1-ar_H)}{(1-ar_H) + \sqrt{(1+ar_H)^2 + 4a^2r_H^2}}$$

From Eq.(36), where $a$ is $\frac{1+\Lambda r_H^2}{r_H}$ or $\frac{16 - c^2 r_H^4 + 16\Lambda r_H^2}{16 r_H}$, and $c$ is a uncertain constant.

The expression of the deflection angle in spherically symmetric spacetime is Eq. (36). With the map of the geometric structure, the lens equation can be simplified as Eq.(37) when the condition $(\beta, \theta \sim 0)$ is satisfied.

$$\beta=\theta-\frac{D_{LS}}{D_{OS}}\Delta\alpha_n, \tag{37}$$

where $\Delta\alpha_n = \alpha - 2n\pi$, $n$ is the number of loops that the photon surround the gravitational fields. Then we can obtain the position of the image $\theta_n$ and the magnification $\mu_n$ of the $n^{th}$ image

$$\theta_n = \theta_n^0 + \frac{u_m e_n (\beta - \theta_n^0) D_{OS}}{\bar{a}\beta D_{LS} D_{OL}}, \tag{38}$$

$$\mu_n = e_n \frac{u_m^2 (1+e_n) D_{OS}}{\bar{a}\beta D_{OL}^2 D_{LS}}. \tag{39}$$

Here,

$$\theta_n^0 = \frac{u_m (1+e_n)}{D_{OL}}, \quad e_n = exp\left(\frac{\bar{b}-2n\pi}{\bar{a}}\right). \tag{40}$$

From the Eq. (40), the approximate expression $\mu_n \sim e^{-n}/(\beta D_{OL}^2)$ indicates that, the brightness of the image decreased with the exponent, and the first level of the image is the brightest level. For the same level, with the distance $D_{OL}$ increasing, the brightness of the image decreased. On the other hand, the smaller the angle $\beta$, the more bright of the image. When the levels $n$ approach to $\infty$, the relationship between the impact parameter and the asymptotic position of the relativistic image is,

$$\theta_\infty = \frac{u_m}{D_{OL}}. \tag{41}$$

For the convenience, we usually separate the outermost image from the other images, which is $\theta_1$, and the surplus images are regarded as a group $\theta_\infty$. So the angular separation is

$$s = \theta_1 - \theta_\infty = \theta_\infty \cdot exp\left(\frac{\bar{b}-2\pi}{\bar{a}}\right). \tag{42}$$

Then, the ratio of luminous flux between the first image and other ones can be expressed as

$$\tilde{r} = \mu_1 \bigg/ \sum_{2}^{\infty} \mu_n = exp(2\pi/\bar{a}). \tag{43}$$

Where, the position of image, the angular separation and the ratio of luminous flux are all observable quantities. With the measure of those observable quantities, the limit coefficient of the strong gravitational lensing and the impact parameter are obtained. Therefore, we can test the validity of the strong gravitational lensing model through the analysis between the observable quantities and the theoretical value.

## 3 Conclusion

Because the particularity of Proca field, the strong gravitational fields in the Einstein-Proca theory are more general than the spherically symmetric spacetime. Hence, we studied the strong gravitational lensing of a spherically symmetric spacetime in the Einstein-Proca theory by Bozza's method. And the limit coefficient, the deflection angle, the position of image, the angular separation and the ratio of luminous flux are obtained. Eqs.(3), (4), (34-36) imply that, the horizon of the black hole does not only effect the Hawking radiation but also influence the strong gravitational lensing in the carved spacetime. Because the horizon $r_H$ is determined by the metric in the gravitational field, the gravitational lensing is related the metric of the black hole closely. As a result, we can use the gravitational lensing as a tool to distinguish the various different gravitational fields.

Furthermore, the gravitational lensing affords a very convenient and effective way to distinguish the different gravitational fields. According to the observable value of the strong gravitational lensing, we can distinguish the gravitational field in Einstein-Proca theory from other gravitational fields. Obviously, the strong gravitational lensing has become the one of the most important research fields in the cosmology and the astronomy. In the future work, we will pay more attention to this issue.

**Acknowledgement**：This work was supported by the National Foundation of china under Grant NO.11178018.